\documentclass[12pt]{article}
\usepackage{amssymb, amsmath}
\usepackage{amsfonts}
\usepackage{color}
\newcommand{\Section}[1]%
{\section{#1}\setcounter{equation}{0}%
\setcounter{theorem}{0}}
\newtheorem{theorem}{Theorem}

\def\ze{\mathbb{Z}}

\newenvironment{proof}[1]%
{\par\noindent{\em #1:\ }}%
{~\rule{2mm}{2mm}\par\bigskip}

\begin{document}
\newpage\thispagestyle{empty}
{\topskip 2cm
\begin{center}
{\Large\bf Bose-Einstein Condensation for Lattice Bosons\\} 
\bigskip\bigskip
{\Large Tohru Koma
\footnote{\small \it Department of Physics, Gakushuin University (retired), Mejiro, Toshima-ku, Tokyo 171-8588, JAPAN}
\\}
\end{center}
\vfil
\noindent
{\bf Abstract:} We present a class of models of interacting lattice bosons which show complete Bose-Einstein 
condensation for the ground state.    
\par
\vfil}

\Section{Introduction}

As is well known, it is very difficult to prove the existence of Bose-Einstein condensation for interacting bosons. 
Actually, the mathematically rigorous results are still rare. (See, e.g., \cite{ALSSY,LSSY} and references therein.) 
In this paper, we present a class of interacting boson models on the $d$-dimensional hypercubic lattice for $d\ge 1$. 
We prove that the models show complete Bose-Einstein condensation for the ground state. 
Although the interactions in the Hamiltonians are not necessarily standard, each of them is finite range or 
rapidly decays with the diameter of the support.    

The present paper is organized as follows: In the next Sec.~\ref{Sec:ModelResults}, we give the precise 
definition of the simplest model in our class and the corresponding theorem with the proof. In Sec.~\ref{Sec:ExtDiscussion}, 
general models in our class are presented, and we also give an argument about the absence of Bose-Einstein condensation \cite{PS} 
for certain models at zero temperature. However, the argument is not necessarily mathematically rigorous.  

\Section{Model and Result}
\label{Sec:ModelResults} 

We begin with the simplest model in our class. 
The extension to general models will be given in the next Sec.~\ref{Sec:ExtDiscussion}. 

We consider a Bose gas on a $d$-dimensional hypercubic lattice $\Lambda\subset\ze^d$ with the dimension $d\ge 1$. 
The Hamiltonian with nearest neighbor hopping and nearest neighbor interactions is given by 
\begin{equation}
\label{ham}
H_g^{(N)}:=\frac{1}{2}\sum_{x,y\in\Lambda:|x-y|=1}(a_x^\dagger-a_y^\dagger)(a_x-a_y)
+g\sum_{x,y\in\Lambda:|x-y|=1}[n_x(n_x-1)+n_xn_y-a_x^\dagger n_x a_y -a_y^\dagger n_x a_x],
\end{equation}
where $a_x^\dagger$ and $a_x$ are, respectively, the creation and annihilation operators at 
the site $x=(x^{(1)},x^{(2)},\ldots,x^{(d)})$ in the finite lattice $\Lambda\subset\ze^d$; 
$n_x:=a_x^\dagger a_x$ is the number operator at the site $x$, and $g\ge 0$ is the coupling constant. 
As usual, the operators, $a_x^\dagger$ and $a_x$, obey the commutation relations,  
$$
[a_x, a_y^\dagger]=\delta_{x,y}, \quad [a_x, a_y]=0, \ \ \mbox{and} \ \ \ [a_x^\dagger, a_y^\dagger]=0. 
$$
We fix the total number of the bosons to $\sum_x n_x=N$, and we impose the periodic boundary condition. 
Then, the Fourier transform of the operator $a_x$ is given by 
$$
\hat{a}_k:=\frac{1}{\sqrt{|\Lambda|}}\sum_{x\in\Lambda}e^{-ikx} a_x, 
$$
where $k:=(k^{(1)},k^{(2)},\ldots,k^{(d)})$ is the momentum, and $kx=\sum_{i=1}^d k^{(i)}x^{(i)}$; 
$|\Lambda|$ denotes the number of the sites in the finite lattice $\Lambda$. The inverse is 
\begin{equation}
\label{ax}
a_x=\frac{1}{\sqrt{|\Lambda|}}\sum_k e^{ikx} \hat{a}_k. 
\end{equation}  
By using this expression, the kinetic term of the Hamiltonian $H_g^{(N)}$ is written 
\begin{equation}
H_0^{(N)}=\sum_k \mathcal{E}_k \hat{a}_k^\dagger \hat{a}_k
\end{equation}
with the standard dispersion relation,  
\begin{equation}
\mathcal{E}_k:=\sum_{i=1}^d (1-\cos k^{(i)}). 
\end{equation} 

The complete Bose-Einstein condensation state is given by 
\begin{equation}
\Phi_{\rm BEC}^{(N)}:=\frac{1}{\sqrt{N!}}(\hat{a}_0^\dagger)^N|0\rangle, 
\end{equation}
where $|0\rangle$ is the vacuum state, i.e., $\hat{a}_k|0\rangle =0$ for all the momenta $k$. 
\bigskip

Our result is: 

\begin{theorem}
\label{mainTheorem}
The complete Bose-Einstein condensation state $\Phi_{\rm BEC}^{(N)}$ is the unique ground state 
of the Hamiltonian $H_g^{(N)}$ of (\ref{ham}) for any coupling constant $g\ge 0$ in any dimensions $d\ge 1$. 
\end{theorem} 
\medskip

\begin{proof}{Proof}
The summand of the interaction term of the Hamiltonian $H_g^{(N)}$ of (\ref{ham}) is written 
\begin{equation}
\label{PositiveIntSummand}
n_x(n_x-1)+n_xn_y-a_x^\dagger n_x a_y -a_y^\dagger n_x a_x
=(a_x^\dagger - a_y^\dagger)a_x^\dagger a_x(a_x-a_y). 
\end{equation}
One notices that this right-hand side is positive. 
Using the expression (\ref{ax}) of the operator $a_x$ in terms of $\hat{a}_k$, one has 
\begin{equation}
a_x-a_y=\frac{1}{\sqrt{|\Lambda|}}\sum_k (e^{ikx}-e^{iky})\hat{a}_k
=\frac{1}{\sqrt{|\Lambda|}}\sum_{k\ne 0} (e^{ikx}-e^{iky})\hat{a}_k.
\end{equation}
Namely, the two contributions of the zero momentum mode $\hat{a}_0$ cancel with each other in the sum. 
Therefore, one has 
\begin{equation}
(a_x-a_y)\Phi_{\rm BEC}^{(N)}=0.
\end{equation}
Combining this with (\ref{PositiveIntSummand}), the summand of the interaction term also annihilates 
the state $\Phi_{\rm BEC}^{(N)}$, i.e., 
\begin{equation}
[n_x(n_x-1)+n_xn_y-a_x^\dagger n_x a_y -a_y^\dagger n_x a_x]\Phi_{\rm BEC}^{(N)}=0. 
\end{equation}
Immediately, 
\begin{equation}
\langle \Phi_{\rm BEC}^{(N)},H_{\rm int}^{(N)}\Phi_{\rm BEC}^{(N)}\rangle =0, 
\end{equation}
where we have written $H_{\rm int}^{(N)}:=H_g^{(N)}-H_0^{(N)}$ for the interaction term of the Hamiltonian $H_g^{(N)}$. 
This yields  
\begin{equation}
\label{zeroExpPhiBECHg}
\langle \Phi_{\rm BEC}^{(N)},H_g^{(N)}\Phi_{\rm BEC}^{(N)}\rangle =0 
\end{equation}
for the total Hamiltonian $H_g^{(N)}$ as well. 
We write $\Phi_{\rm GS}^{(N)}$ for the ground state of $H_g^{(N)}$. Then, from this result (\ref{zeroExpPhiBECHg}), we have 
\begin{equation}
0=\langle \Phi_{\rm BEC}^{(N)},H_g^{(N)}\Phi_{\rm BEC}^{(N)}\rangle
\ge \langle \Phi_{\rm GS}^{(N)}, H_g^{(N)}\Phi_{\rm GS}^{(N)}\rangle \ge 
\sum_k \mathcal{E}_k \langle \Phi_{\rm GS}^{(N)},\hat{a}_k^\dagger \hat{a}_k\Phi_{\rm GS}^{(N)}\rangle \ge 0, 
\end{equation}
where we have used the positivity $H_{\rm int}^{(N)}\ge 0$ which is obtained from the positivity (\ref{PositiveIntSummand}). 
This implies 
$$ 
\langle \Phi_{\rm GS}^{(N)},\hat{a}_k^\dagger \hat{a}_k\Phi_{\rm GS}^{(N)}\rangle=0
$$
for all $k\ne 0$. Thus, we obtain $\Phi_{\rm GS}^{(N)}=\Phi_{\rm BEC}^{(N)}$, and the ground state is unique. 
\end{proof}
\bigskip

\noindent
{\bf Remark:} The method in the above proof is similar to that in \cite{KKMTT}. 
Namely, both of kinetic and interaction terms in a Hamiltonian annihilate a state. 

\Section{Extension to a Class of Models and Discussion} 
\label{Sec:ExtDiscussion}

In this section, we extend the model of (\ref{ham}) and Theorem~\ref{mainTheorem} to general models. 
We also discuss certain interactions for which the absence of Bose-Einstein condensation can be expected 
at zero temperature. 

A straightforward extension of (\ref{PositiveIntSummand}) is as follows: Consider an interaction Hamiltonian, 
\begin{equation}
\label{Hintextend}
H_{\rm int}^{(N)}=g\sum_{x\in \Lambda} C_x^\dagger A_x C_x, 
\end{equation}
where $A_x$ is a positive local operator and $C_x$ is a local operator which is written in the form, 
\begin{equation}
C_x=\sum_j \alpha_j a_{x_j}, 
\end{equation}
with coefficients $\alpha_j$ which satisfy $\sum_j \alpha_j=0$. 

We can also choose $C_x$ to be  
\begin{equation}
\label{CxK}
C_x=\sum_{y\in \Lambda}K(x-y)a_y,
\end{equation}
where the function $K(x-y)$ is given by the Fourier transform of $\hat{K}(k)$, i.e., 
\begin{equation}
\label{functionK}
K(x-y)=\frac{1}{|\Lambda|}\sum_k \hat{K}(k)e^{ik(x-y)}.
\end{equation}
When the function $\hat{K}(k)$ is sufficiently many times differentiable and satisfies $\hat{K}(0)=0$, 
the operator $C_x$ is local and annihilates the state $\Phi_{\rm BEC}^{(N)}$. Actually, one has 
\begin{equation}
C_x=\frac{1}{\sqrt{|\Lambda|}}\sum_k \hat{K}(k)\hat{a}_ke^{ikx}.
\end{equation} 

Conversely, when the function $\hat{K}(k)$ has a narrow distribution with a high peak at $k=0$, 
we can expect the absence of Bose-Einstein condensation for the zero momentum mode. 
In fact, one has 
\begin{equation}
C_x=\frac{1}{ \sqrt{|\Lambda|} }
\hat{K}(0)\hat{a}_0+\frac{1}{\sqrt{|\Lambda|}}\sum_{k\ne 0}\hat{K}(k)\hat{a}_ke^{ikx}.
\end{equation}
If we choose $A_x=a_x^\dagger a_x$ for the operator $A_x$ in (\ref{Hintextend}), 
then the contribution of the zero momentum mode in (\ref{Hintextend}) is written 
\begin{equation}
\sum_x \frac{1}{|\Lambda|}(\hat{K}(0))^2 \hat{a}_0^\dagger A_x\hat{a}_0
=\sum_x \frac{1}{|\Lambda|}(\hat{K}(0))^2 \hat{a}_0^\dagger a_x^\dagger a_x\hat{a}_0
=\frac{N}{|\Lambda|}(\hat{K}(0))^2 \hat{a}_0^\dagger \hat{a}_0,
\end{equation}
where we have used $\sum_x a_x^\dagger a_x=N$. This right-hand side raises the energy by order 
of the volume $|\Lambda|$ with the large amplitude $(\hat{K}(0))^2$ 
for a fixed particle density $N/|\Lambda|$ when the zero momentum mode exhibits a long-range order.  
Therefore, the long-range order of the zero momentum mode is expected to be suppressed for a large $\hat{K}(0)>0$. 

When $\hat{K}(k)=1$ for all $k$, one has $K(x-y)=\delta_{x,y}$ from (\ref{functionK}). Then, we have   
\begin{equation}
H_{\rm int}^{(N)}=g\sum_{x\in \Lambda} n_x(n_x-1)
\end{equation}
for $A_x=a_x^\dagger a_x$, {from} (\ref{Hintextend}) and (\ref{CxK}). This is one of the standard interactions \cite{Tasakibook}. 
However, it seems to be very difficult to prove the existence of Bose-Einstein condensation in this case 
because the function $\hat{K}(k)=1$ is an intermediate function in the above class.  

Thus, in general, whether Bose-Einstein condensation occurs or not is expected to depend on interactions of models. 
In particular, it may depend on the dimension \cite{PS} of the lattice.


\bigskip\bigskip\bigskip

\noindent
{\bf Acknowledgements:} I would like to thank Akinori Tanaka for helpful comments. 

\end{document}